\documentclass[a4paper,11pt]{article}
\pdfoutput=1 

\usepackage{jheppub} 

\usepackage[T1]{fontenc} 

\title{\boldmath Low-diffusion Xe-He gas mixtures for rare-event detection: Electroluminescence Yield}

\collaboration{The NEXT Collaboration}
\author[13]{A.F.M.~Fernandes,}
\author[13]{C.A.O.~Henriques,}
\author[13]{R.D.P.~Mano,}
\author[21]{D.~Gonz\'alez-D\'iaz,}
\author[4]{C.D.R~Azevedo,}
\author[13]{P.A.O.C.~Silva,}
\author[16,9,a]{J.J.~G\'omez-Cadenas,\note[a]{NEXT Co-spokesperson.}}
\author[13]{E.D.C.~Freitas,}
\author[13]{L.M.P.~Fernandes,}
\author[13,b]{C.M.B.~Monteiro, \note[b]{Corresponding author.}}
\author[2]{C.~Adams,}
\author[19]{V.~\'Alvarez,}
\author[6]{L.~Arazi,}
\author[20]{I.J.~Arnquist,}
\author[2]{K.~Bailey,}
\author[22]{F.~Ballester,}
\author[19]{J.M.~Benlloch-Rodr\'{i}guez,}
\author[14]{F.I.G.M.~Borges,}
\author[3]{N.~Byrnes,}
\author[19]{S.~C\'arcel,}
\author[19]{J.V.~Carri\'on,}
\author[23]{S.~Cebri\'an,}
\author[20]{E.~Church,}
\author[14]{C.A.N.~Conde,}
\author[11]{T.~Contreras,}
\author[21,16]{G.~D\'iaz,}
\author[19]{J.~D\'iaz,}
\author[5]{M.~Diesburg,}
\author[14]{J.~Escada,}
\author[22]{R.~Esteve,}
\author[19]{R.~Felkai,}
\author[16,9]{P.~Ferrario,}
\author[4]{A.L.~Ferreira,}
\author[16]{J.~Generowicz,}
\author[11]{S.~Ghosh,}
\author[8]{A.~Goldschmidt,}
\author[11]{R.~Guenette,}
\author[10]{R.M.~Guti\'errez,}
\author[11]{J.~Haefner,}
\author[2]{K.~Hafidi,}
\author[1]{J.~Hauptman,}
\author[21]{J.A.~Hernando~Morata,}
\author[16,19]{P.~Herrero,}
\author[22]{V.~Herrero,}
\author[6,7]{Y.~Ifergan,}
\author[2]{S.~Johnston,}
\author[3]{B.J.P.~Jones,}
\author[19]{M.~Kekic,}
\author[18]{L.~Labarga,}
\author[3]{A.~Laing,}
\author[5]{P.~Lebrun,}
\author[19]{N.~L\'opez-March,}
\author[10]{M.~Losada,}
\author[11]{J.~Mart\'in-Albo,}
\author[16]{A.~Mart\'inez,}
\author[19,21,c]{G.~Mart\'inez-Lema,\note[c]{Now at Weizmann Institute of Science, Israel.}}
\author[3]{A.D.~McDonald,}
\author[16]{F.~Monrabal,}
\author[22]{F.J.~Mora,}
\author[19]{J.~Mu\~noz Vidal,}
\author[19]{P.~Novella,}
\author[3,a]{D.R.~Nygren,}
\author[19]{B.~Palmeiro,}
\author[5]{A.~Para,}
\author[12]{J.~P\'erez,}
\author[3]{F.~Psihas,}
\author[19]{M.~Querol,}
\author[19]{J.~Renner,}
\author[2]{J.~Repond,}
\author[2]{S.~Riordan,}
\author[17]{L.~Ripoll,}
\author[10]{Y.~Rodr\'iguez Garc\'ia,}
\author[22]{J.~Rodr\'iguez,}
\author[3]{L.~Rogers,}
\author[16,12]{B.~Romeo,}
\author[19]{C.~Romo-Luque,}
\author[14]{F.P.~Santos,}
\author[13]{J.M.F. dos~Santos,}
\author[6]{A.~Sim\'on,}
\author[15,d]{C.~Sofka,\note[d]{Now at University of Texas at Austin, USA.}}
\author[19]{M.~Sorel,}
\author[15]{T.~Stiegler,}
\author[22]{J.F.~Toledo,}
\author[16]{J.~Torrent,}
\author[19]{A.~Us\'on,}
\author[4]{J.F.C.A.~Veloso,}
\author[15]{R.~Webb,}
\author[6,e]{R.~Weiss-Babai,\note[e]{On leave from Soreq Nuclear Research Center, Yavneh, Israel.}}
\author[15,f]{J.T.~White,\note[f]{Deceased.}}
\author[3]{K.~Woodruff,}
\author[19]{N.~Yahlali}
\emailAdd{cristinam@uc.pt}
\affiliation[1]{
Department of Physics and Astronomy, Iowa State University, 12 Physics Hall, Ames, IA 50011-3160, USA}
\affiliation[2]{
Argonne National Laboratory, Argonne, IL 60439, USA}
\affiliation[3]{
Department of Physics, University of Texas at Arlington, Arlington, TX 76019, USA}
\affiliation[4]{
Institute of Nanostructures, Nanomodelling and Nanofabrication (i3N), Universidade de Aveiro, Campus de Santiago, Aveiro, 3810-193, Portugal}
\affiliation[5]{
Fermi National Accelerator Laboratory, Batavia, IL 60510, USA}
\affiliation[6]{
Nuclear Engineering Unit, Faculty of Engineering Sciences, Ben-Gurion University of the Negev, P.O.B. 653, Beer-Sheva, 8410501, Israel}
\affiliation[7]{
Nuclear Research Center Negev, Beer-Sheva, 84190, Israel}
\affiliation[8]{
Lawrence Berkeley National Laboratory (LBNL), 1 Cyclotron Road, Berkeley, CA 94720, USA}
\affiliation[9]{
Ikerbasque, Basque Foundation for Science, Bilbao, E-48013, Spain}
\affiliation[10]{
Centro de Investigaci\'on en Ciencias B\'asicas y Aplicadas, Universidad Antonio Nari\~no, Sede Circunvalar, Carretera 3 Este No.\ 47 A-15, Bogot\'a, Colombia}
\affiliation[11]{
Department of Physics, Harvard University, Cambridge, MA 02138, USA}
\affiliation[12]{
Laboratorio Subterr\'aneo de Canfranc, Paseo de los Ayerbe s/n, Canfranc Estaci\'on, E-22880, Spain}
\affiliation[13]{
LIBPhys, Physics Department, University of Coimbra, Rua Larga, Coimbra, 3004-516, Portugal}
\affiliation[14]{
LIP, Department of Physics, University of Coimbra, Coimbra, 3004-516, Portugal}
\affiliation[15]{
Department of Physics and Astronomy, Texas A\&M University, College Station, TX 77843-4242, USA}
\affiliation[16]{
Donostia International Physics Center (DIPC), Paseo Manuel Lardizabal, 4, Donostia-San Sebastian, E-20018, Spain}
\affiliation[17]{
Escola Polit\`ecnica Superior, Universitat de Girona, Av.~Montilivi, s/n, Girona, E-17071, Spain}
\affiliation[18]{
Departamento de F\'isica Te\'orica, Universidad Aut\'onoma de Madrid, Campus de Cantoblanco, Madrid, E-28049, Spain}
\affiliation[19]{
Instituto de F\'isica Corpuscular (IFIC), CSIC \& Universitat de Val\`encia, Calle Catedr\'atico Jos\'e Beltr\'an, 2, Paterna, E-46980, Spain}
\affiliation[20]{
Pacific Northwest National Laboratory (PNNL), Richland, WA 99352, USA}
\affiliation[21]{
Instituto Gallego de F\'isica de Altas Energ\'ias, Univ.\ de Santiago de Compostela, Campus sur, R\'ua Xos\'e Mar\'ia Su\'arez N\'u\~nez, s/n, Santiago de Compostela, E-15782, Spain}
\affiliation[22]{
Instituto de Instrumentaci\'on para Imagen Molecular (I3M), Centro Mixto CSIC - Universitat Polit\`ecnica de Val\`encia, Camino de Vera s/n, Valencia, E-46022, Spain}
\affiliation[23]{
Laboratorio de F\'isica Nuclear y Astropart\'iculas, Universidad de Zaragoza, Calle Pedro Cerbuna, 12, Zaragoza, E-50009, Spain}

\abstract{High pressure xenon Time Projection Chambers (TPC) based on secondary scintillation (electroluminescence) signal amplification are being proposed for rare event detection such as directional dark matter, double electron capture and double beta decay detection. The discrimination of the rare event through the topological signature of primary ionisation trails is a major asset for this type of TPC when compared to single liquid or double-phase TPCs, limited mainly by the high electron diffusion in pure xenon. Helium admixtures with xenon can be an attractive solution to reduce the electron diffusion significantly, improving the discrimination efficiency of these optical TPCs. We have measured the electroluminescence (EL) yield of Xe–He mixtures, in the range of 0 to 30$\%$ He and demonstrated the small impact on the EL yield of the addition of helium to pure xenon. For a typical reduced electric field of 2.5 kV/cm/bar in the scintillation region, the EL yield is lowered by $ \sim 2 \%$, $3 \%$, $6 \%$ and $10 \%$ for $10\%$, $15\%$, $20\%$  and $30\%$ of helium concentration, respectively. This decrease is less than what has been obtained from the most recent simulation framework in the literature. The impact of the addition of helium on EL statistical fluctuations is negligible, within the experimental uncertainties. The present results are an important benchmark for the simulation tools to be applied to future optical TPCs based on Xe-He mixtures.}

\begin{document} 
\maketitle
\flushbottom

\section{Introduction}
\label{sec:intro}
The nature of Dark Matter and Neutrinos, either Majorana or Dirac, is of major importance for human knowledge, at present. To address these issues, optical TPCs are being proposed and/or developed for rare event detection, such as directional dark matter~\cite{1,2,3} and double beta decay (DBD) detection~\cite{4,5}. In addition, they are potential candidates for double electron capture (DEC) detection, substituting for proportional counters~\cite{6,7,8}. Many of these implementations involve operation in high pressure xenon.

The amplification of ionisation electron signals through xenon electroluminescence (EL) allows achieving both higher detector signal-to-noise ratio~\cite{9,10}, due to the additional gain of the photosensor, and lower statistical fluctuations when compared to charge avalanche multiplication~\cite{11}. At 10 bar, the best energy resolution achieved with a 1kg-scale prototype based on Micromegas was extrapolated to around 3$\%$-FWHM at the xenon $Q_{\beta \beta}$ (2.45 MeV) ~\cite{12}, while a 1kg- and a 10 kg-scale EL-based TPC achieved energy resolution values consistently below 1$\%$-FWHM ~\cite{13,14}. The EL readout through photosensors electrically and mechanically decouples the amplification region from the readout, rendering the system more immune to electronic noise, radiofrequency pickup and high voltage issues. When compared to LXe-based TPCs, event detection in the gas phase achieves a better energy resolution and allows for discrimination of the rare event through its topological signature, as demonstrated for double electron capture and double beta decay detection ~\cite{7,8,12,15,16,17,18,19}. The reduced dimensions of the ionisation trace in LXe rules out any topology-based pattern recognition for events of few MeV or below.

The NEXT collaboration aims at the detection of neutrinoless double beta decay in \textsuperscript{136}Xe~\cite{4} and, presently, operates the largest HPXe optical-TPC, based on EL for ionisation signal amplification~\cite{19}. The unambiguous observation of this decay would demonstrate leptonic number violation and prove the Majorana nature of the neutrino, presenting a breakthrough for new physics, beyond the Standard Model.

\begin{figure}[tbp]
\centering 
\includegraphics[]{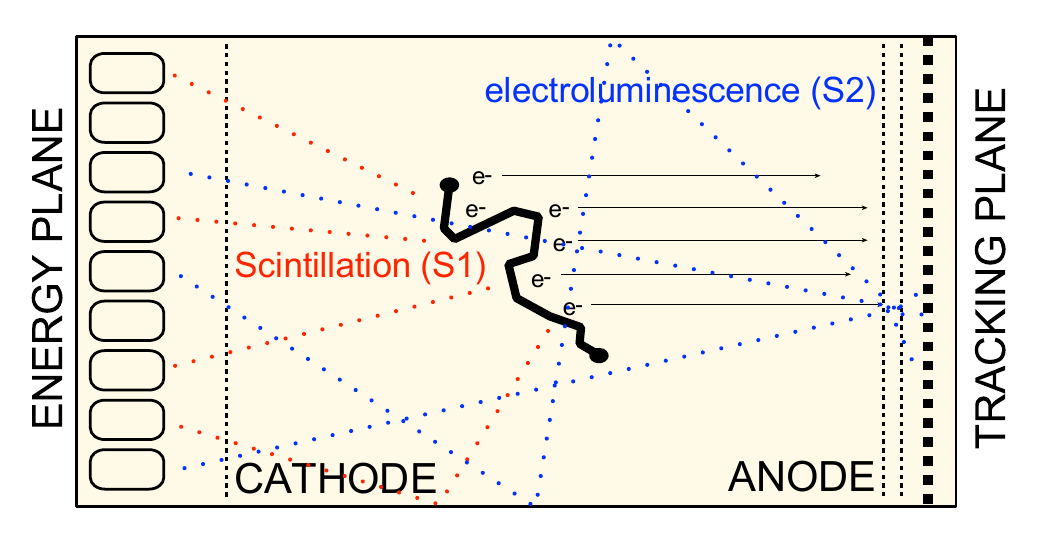}
\caption{\label{fig:next}Schematic of the EL-based TPC developed by the NEXT collaboration for double-beta decay searches in \textsuperscript{136}Xe~\cite{20}.}
\end{figure}

The schematic of a typical optical TPC, as the one that has been developed by the NEXT collaboration, is presented in figure~\ref{fig:next}. The radiation interaction takes place in the conversion (drift) region, the sensitive volume, exciting and/or ionising the gas atoms/molecules and leading to the emission of primary scintillation (providing the t\textsubscript{0} signal of the event, i.e. the start-of-event time-stamp) from the gas de-excitation and, in the case of highly ionising particles, also from electron/ion recombination. An electric field of intensity below the gas excitation threshold is applied to this region to minimize recombination and to guide the primary electrons towards the scintillation region. The scintillation region is defined by two parallel electrodes, being the electric field intensity set between the gas excitation and the gas ionisation thresholds. Upon crossing this region, each electron attains, from the electric field, enough kinetic energy to excite but not ionise the gas atoms/molecules, by electron impact, leading to high scintillation-output (electroluminescence) ensuing the gas deexcitation processes, without charge avalanche formation. The x- and y-positions of the primary electrons arriving at the EL region are determined by reading out the EL by means of a pixelated plane of photosensors while, from the difference in time between the primary and the EL scintillation pulses, the z-position at which the ionisation event took place can be determined. 

EL yields for xenon and argon have been measured for uniform electric fields ~\cite{21,22,23,24}, as well as for electron avalanches produced in modern micropatterned electron multipliers, e.g. GEM, THGEM, MHSP and Micromegas ~\cite{25,26,27}. However, it is to be noted that the excellent energy resolution that can be obtained with the EL readout, e.g., needed for efficient background discrimination in neutrinoless DBD detection, is only reached for the EL produced in electric fields of values that are below or near the onset of electron multiplication. The statistical fluctuations in the EL produced at electric fields below the onset of electron multiplication are negligible when compared to those associated with the primary ionisation formation, while the statistical fluctuations of the EL produced in electron avalanches are dominated by the much larger variance of the total number of electrons produced in the avalanches ~\cite{11,28}.

Although the topological signature capability of HPXe TPCs based on EL has been demonstrated, e.g. ~\cite{16,17,18}, the large electron diffusion in pure xenon presents a limitation, particularly for large drift distances. Diffusion hinders the finer details of the ionisation trail, and the discrimination based on the topological signature becomes less effective ~\cite{29}. For the low electric field values (few tens of V/cm/bar) used in the NEXT TPC, electron transverse diffusion may be as high as 10 mm/$\sqrt{\textrm{m}}$, making the pattern recognition of the primary ionisation trail difficult at the 1-m drift scale ~\cite{29}. Recent studies have demonstrated that the addition of molecular gases, such as CO\textsubscript{2}, CH\textsubscript{4} and CF\textsubscript{4}, to pure xenon, at sub-percent concentration levels, reduces the electron diffusion to the level of $ \sim $2 mm/$\sqrt{\textrm{m}}$, without jeopardizing the performance of the TPC in terms of EL yield and energy resolution, with CH\textsubscript{4} being the most suitable candidate ~\cite{20,30,31}.

On the other hand, one has to take into account that standard xenon purification through hot getters may not be suitable for the chosen molecular additive, or else, the getter operating temperature may have to be lowered to prevent molecular breakdown, which may affect the gas cleaning efficiency. In addition, the cryogenic separation of the molecular additive has to be made efficiently enough in order to prevent any loss of the expensive, enriched xenon. CH\textsubscript{4}, at the same time, presents some degree of excimer-quenching ~\cite{30,31}, which could limit the primary scintillation yield and, therefore, the calibration for low-energy events.

While the aforementioned aspects are yet to be studied in higher detail in real-size detectors, and may be certainly overcome, the addition of a noble gas such as He could offer an alternative solution, free from those limitations~\cite{32,33}. Simulation studies of electron drift parameters, as well as primary and secondary scintillation yields of Xe-He mixtures have been carried out recently~\cite{33}. The significantly lower mass of helium atoms, when compared to xenon, allows more efficient cooling of the electrons along the drift path. The result of the simulation studies indicate that a transverse diffusion of 2.5 mm/$\sqrt{\textrm{m}}$ is achievable with 15$\%$  of helium additive without a significant degradation of the intrinsic energy resolution and of the EL-yield. 

The advantages of using helium as additive would be of utmost impact as Xe-He mixtures would share exactly the same purification system as pure xenon and full xenon cryogenic recovery would be much easier. Yet, the use of such mixture would reduce the amount of the source isotope in the detector. The final value of the helium concentration should be a compromise between an improvement of the background rejection factor and a reduction of the active mass that is needed to maximize sensitivity, as noted in~\cite{33}.

Experimental studies for the electron drift parameters in Xe-He mixtures have been carried out very recently~\cite{34}. The impact of helium on the electron diffusion was not as substantial as anticipated, especially in the region corresponding to the Ramsauer minimum (around 10V/cm/bar for pure xenon and 25V/cm/bar for 15$\%$  He admixture) but remained in agreement with simulations outside that region. On the other hand, the impact of the helium additive on the xenon EL yield had yet to be determined experimentally in order to understand the scope of the use of these mixtures in EL-TPCs. 

Following the electroluminescence studies on Xe mixtures with sub-percent concentration levels of CO\textsubscript{2}, CH\textsubscript{4} and CF\textsubscript{4}~\cite{31}, in the present work we study the EL yield of Xe-He mixtures, in the range from 0 to 30$\%$  of helium and the impact of the helium addition on the TPC energy resolution. 

\section{Experimental setup}
\label{sec:exp}
\begin{figure}[tbp]
\centering 
\includegraphics[width=13cm]{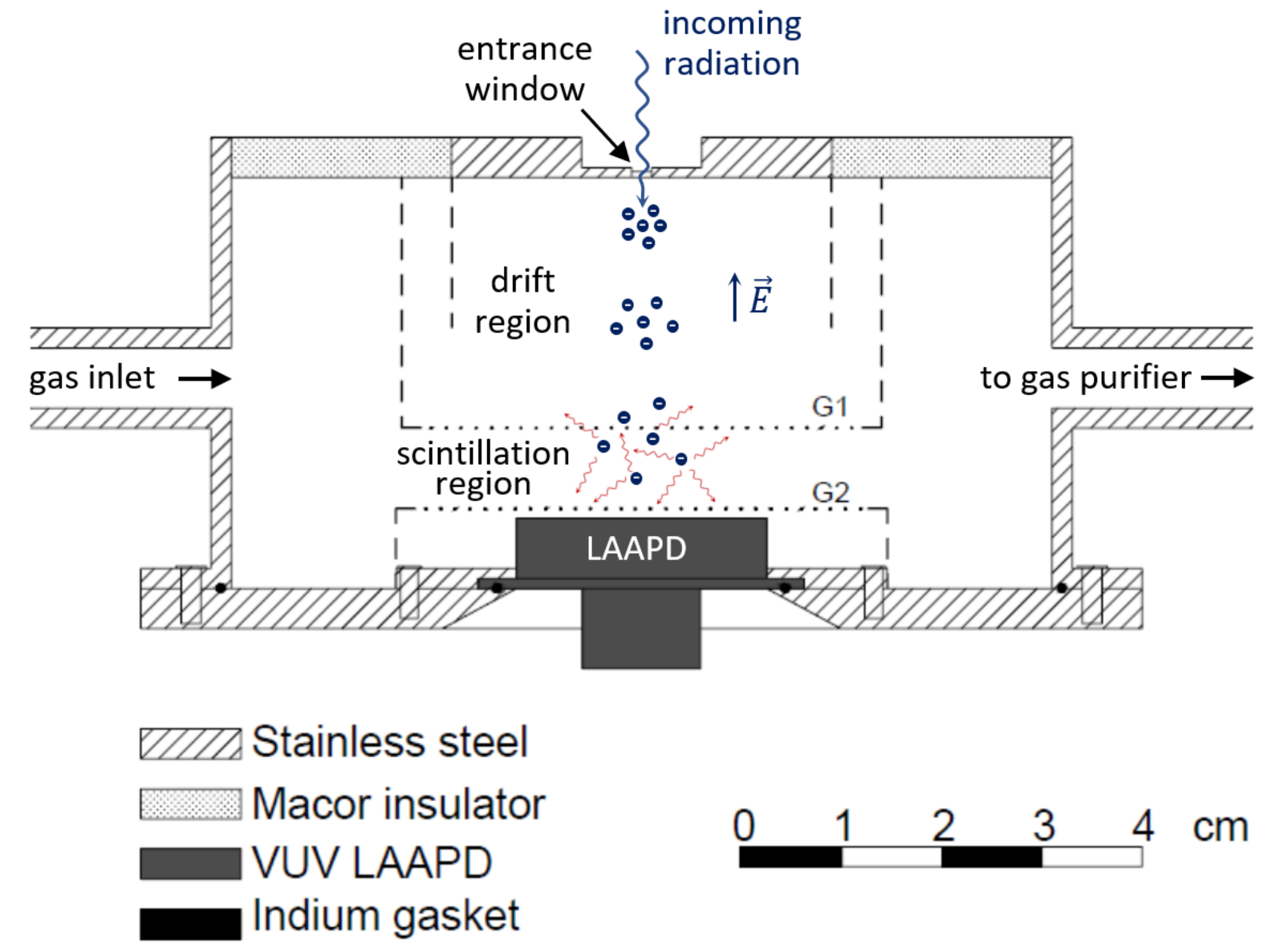}
\caption{\label{fig:scheme}Schematic of the Gas Proportional Scintillation Counter used for this work. A VUV-sensitive LAAPD is used as photosensor and the gas is continuously purified, circulating by convection through SAES St707 getters.}
\end{figure}

\begin{figure}[tbp]
\centering 
\includegraphics[]{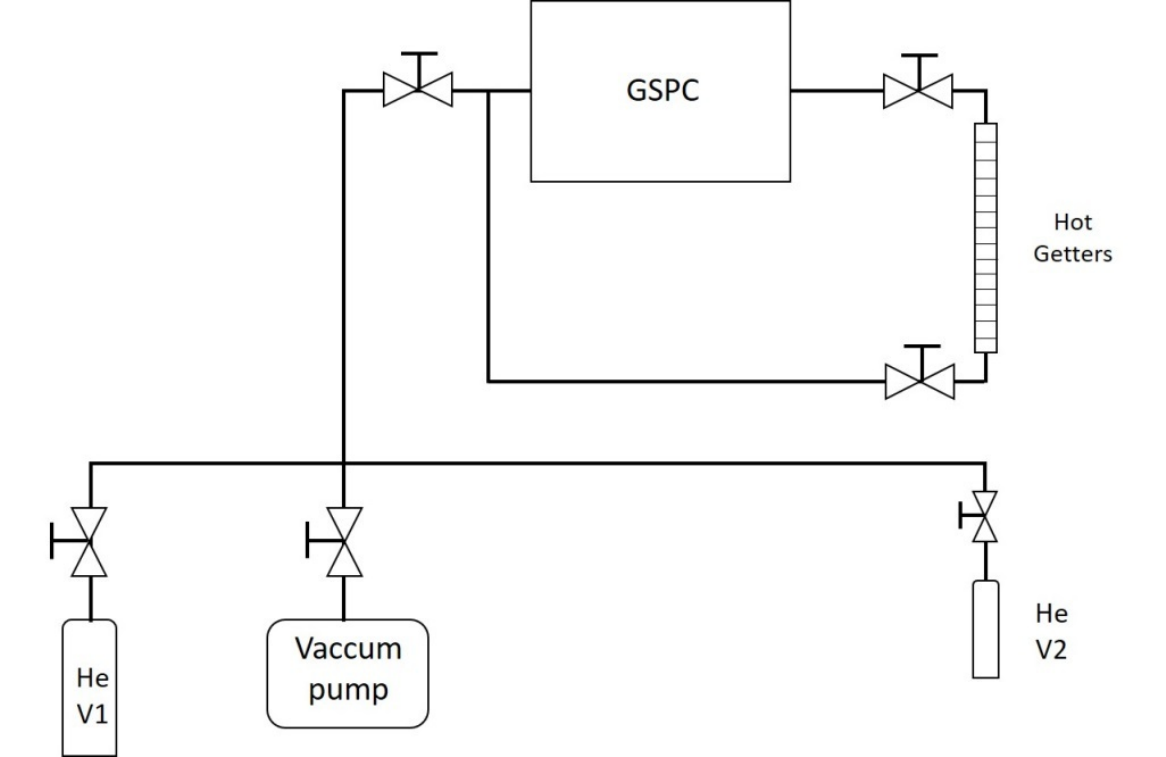}
\caption{\label{fig:GPSC}Layout of the experimental setup, including the GPSC, the gas circulation and purifying system with SAES St-707 getters and the two helium volumes for implementing the admixtures.}
\end{figure}

The EL studies were performed in a Gas Proportional Scintillation Counter (GPSC) that has a large area avalanche photodiode (LAAPD) inside for the EL readout. The GPSC is depicted schematically in figure~\ref{fig:scheme} and had already been used in~\cite{35,36} with pure xenon and pure argon filling, respectively. It has a 2.5-cm deep conversion/drift region and a 0.8-cm deep scintillation region. The GPSC was filled at pressures around 1.2 bar, with the gas being continuously purified through St707 SAES getters that were kept at 150ºC, and circulated by convection. The gas circulation and purifying system is a ``U''-tubing that closes up in the GPSC gas-in and -out connections, the getters placed inside one of its vertical arms, figure~\ref{fig:GPSC}. 

Grids G1 and G2 are of highly transparent stainless steel wire, 80-$ \mu $m in diameter and 900-$ \mu $m spacing, delimiting the scintillation region. The detector radiation window is made of Melinex, 6-$ \mu $m thick, 2 mm in diameter. A Macor piece isolates the holders of both radiation window and grid G1 and is vacuum-sealed to the stainless steel with a low vapour pressure epoxy. The LAAPD is vacuum-sealed by compressing the photodiode enclosure against the stainless steel detector body using an indium ring.

The LAAPD enclosure and G2 are kept at ground potential. Therefore, the electric field in the scintillation region is set by the voltage of G1, while the electric field in the drift region is set by the voltage difference between the GPSC radiation window and G1. The LAAPD is an Advanced Photonics Inc., deep-UV enhanced series~\cite{37} with a 16-mm active diameter. Throughout the measurements, the LAAPD bias was kept at 1840 V, corresponding to a gain of $\sim$150. The response of the LAAPD to the xenon VUV EL can be found in detail in~\cite{38}. A 1-mm collimated 5.9-keV x-ray beam from a \textsuperscript{55}Fe radioactive source, filtered with a chromium film, was used to irradiate the GPSC along its axis. The LAAPD signals were fed through a low-noise, 1.5 V/pC, charge pre-amplifier to an amplifier with 2-$ \mu $s shaping time, and were pulse-height analysed with a multi-channel analyser.

Two small volumes, with well-established volume ratios, were connected to the GPSC through vacuum valves, figure~\ref{fig:GPSC}. The whole system was pumped down to pressures in the 2x10\textsuperscript{-6} mbar range for several hours; the volumes were filled with the proper amount of He, previously calculated to obtain the intended Xe-He concentrations, and the GPSC was, afterwards, filled with pure xenon. Therefore, in a single run, the EL output of the GPSC was studied for pure xenon and for two different helium concentrations, without the need to switch off the GPSC and LAAPD bias voltages, and the GPSC response to the 5.9-keV x-rays was continuously monitored, while the study of a given mixture was in progress. The xenon gas purity was of grade 4.8 from Messer while helium was of grade 5.0. A relative uncertainty below 4$\%$ in the He concentration results from both the pressure gauge precision and the uncertainty in the ratio of the volumes.

\section{Method}
\label{sec:method}

\begin{figure}[tbp]
\centering 
\includegraphics[width=12.0cm]{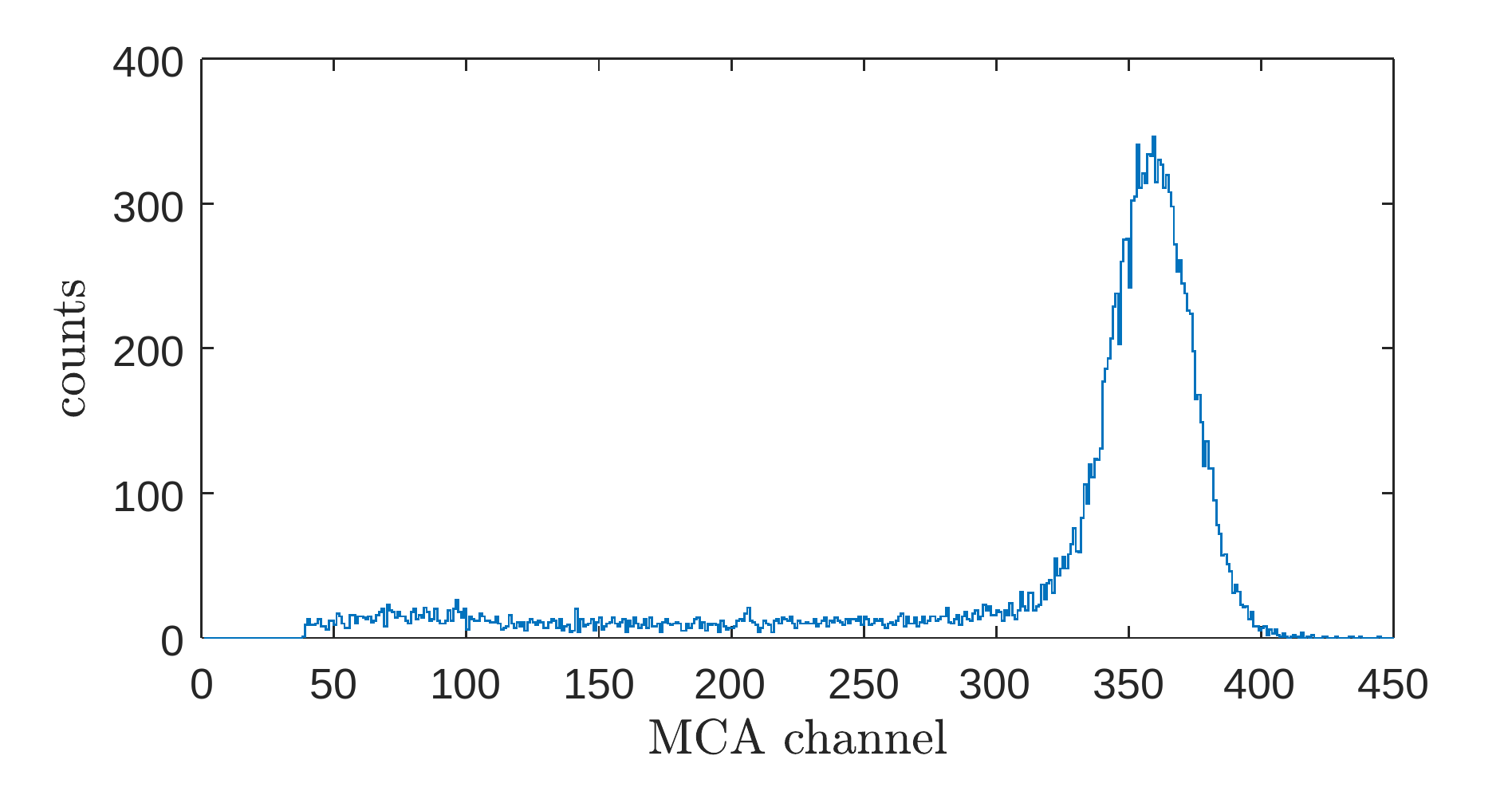}
\caption{\label{fig:fe_spectrum}Pulse-height distribution for 5.9-keV x-rays absorbed in the GPSC active volume filled with Xe-15$\%$ He, for a reduced electric field of 2.4 kV/cm/bar.}
\end{figure}

Figure~\ref{fig:fe_spectrum} depicts a typical response of the GPSC to 5.9-keV x-rays. The primary scintillation produced by x-ray interaction is more than 3 orders of magnitude lower than the EL output ~\cite{39} and, thus, is well within the electronic noise. Nevertheless, the primary scintillation can be measured by averaging a significant number, of the order of several thousand, of waveforms, triggering on the EL using a constant trigger level. For that purpose, we have used a LeCroy WaveRunner 610Zi digital oscilloscope. More detailed information on this GPSC’s response to x-rays can be found in~\cite{35,36}.

The full-absorption peaks were fitted to Gaussian functions, superimposed on a linear background, from which the centroid, taken as the pulse amplitude, and the FWHM were determined. For each helium concentration, we have studied the centroid position of the full-absorption peak and its relative FWHM, the GPSC energy resolution, as a function of reduced electric field E/p, the electric field divided by the gas pressure, in the scintillation region. The reduced electric field in the drift region was kept below the gas excitation threshold. 

In this work, only relative values were measured for the EL yield. In each run, absolute values for the reduced EL yield, Y/p, were obtained by normalizing the pulse amplitude measured for pure xenon at an E/p of $ \sim$2.0 kV/cm/bar to the corresponding absolute value obtained in~\cite{33}. The same normalization constant has been used, then, to normalize the remaining centroid values obtained for the different E/p for pure xenon and for the two Xe-He mixtures studied in that run. Small variations that may eventually occur in the LAAPD leak current during a run were taken into account to correct the centroid values obtained along that same run, being those corrections at the level of less than a few percent.

\section{Experimental results and discussion}
\label{sec:exp_disc}
\begin{figure}[tbp]
\centering 
\includegraphics[]{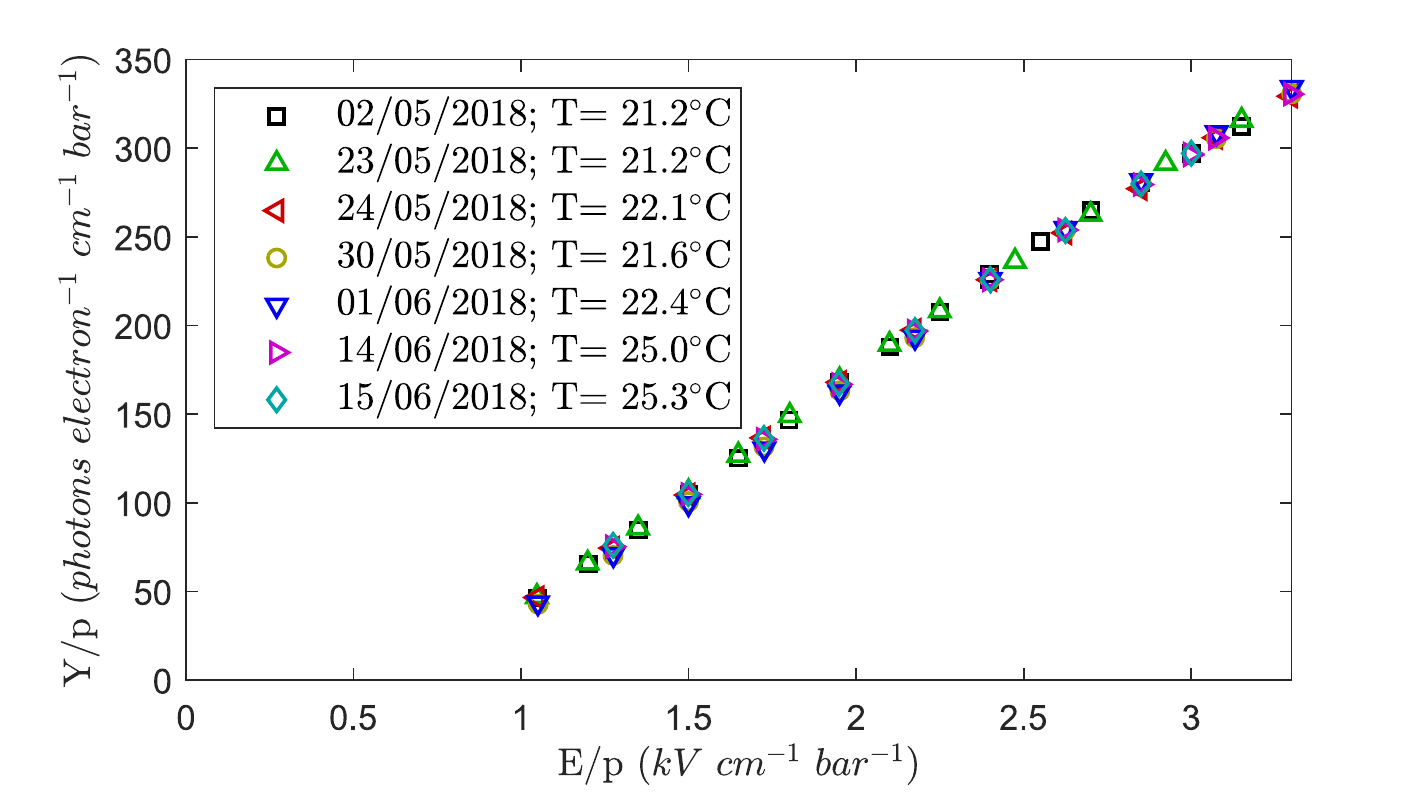}
\caption{\label{fig:yield}EL yield, Y/p, for pure xenon, as a function of reduced electric field E/p applied to the scintillation region, obtained in different runs with different LAAPD operation conditions.}
\end{figure}

The consistency of our experimental procedure is shown in figure~\ref{fig:yield}, where the reduced EL yield (Y/p) is depicted as a function of reduced electric field (E/p) applied to the scintillation region for pure xenon. The different data sets were taken at different times along the whole experimental campaign and have different operation conditions such as the LAAPD temperature, leak current and gain. A good reproducibility of the normalized experimental results is observed. From the data of figure~\ref{fig:yield} we determined the values for the amplification parameter for EL to be 136 $ \pm $ 1 photons/kV, the slope of the linear fit. The average scintillation threshold for EL, the linear fit interception with the horizontal axis, is 0.69 $ \pm $  0.04 kV/cm/bar. This value is in good agreement with both the simulation studies and the experimental values presented in the literature~\cite{21,23,30,40}. From the energy resolution data, an intrinsic energy resolution around 6.4$\%$  FWHM and a Fano factor of 0.20 $ \pm $  0.04 were estimated. The latter value is similar to that estimated in a driftless Xe-GPSC~\cite{20,31} and is in good agreement with the values reported in the literature~\cite{41,42,43,44}.

\begin{figure}[tbp]
\centering 
\includegraphics[width=10cm]{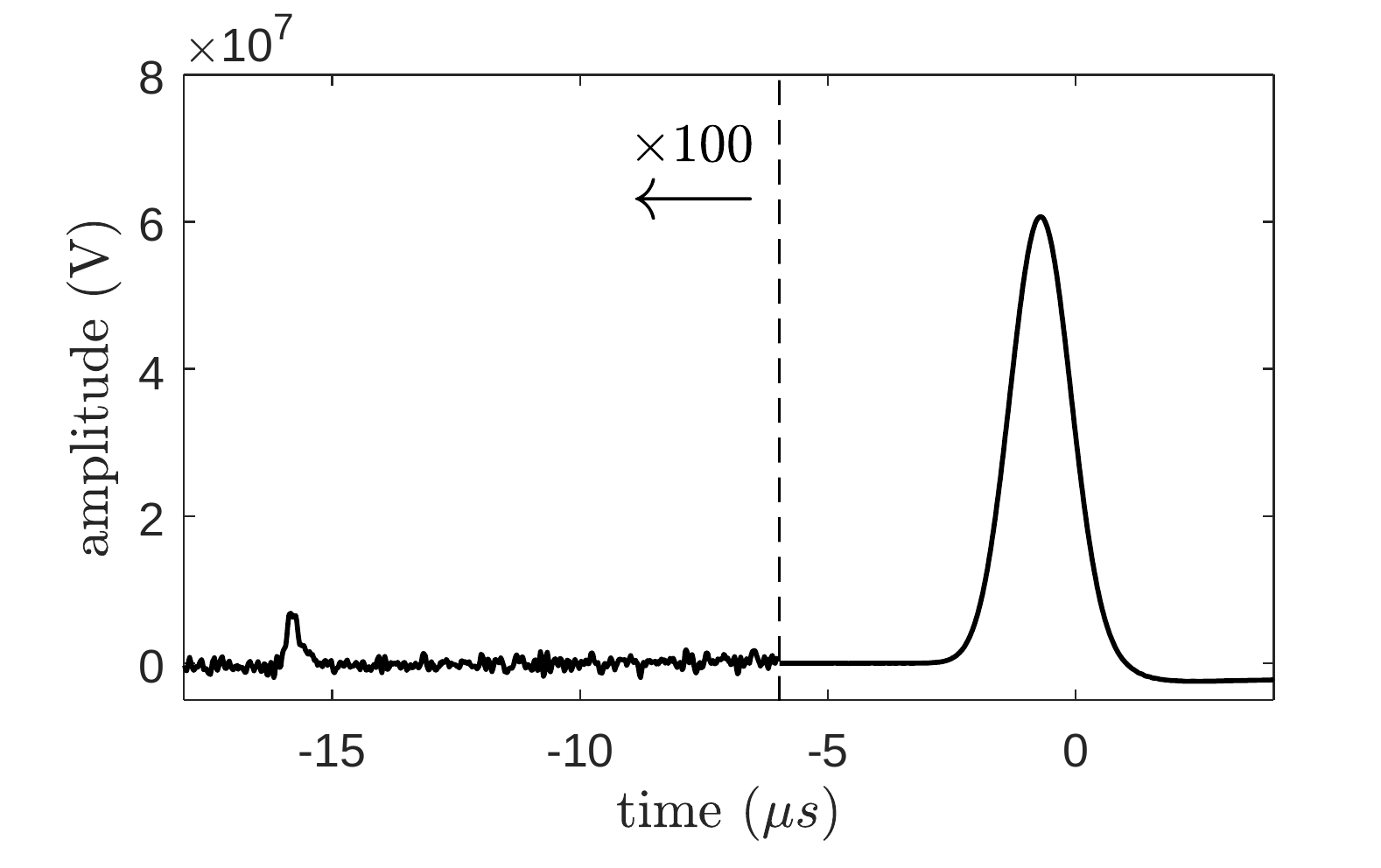}
\caption{\label{fig:waveform}Typical average waveform produced by alpha particles (a few tens of Hz) in pure xenon at 1.1 bar, for a mean reduced electric field of 376 V/cm/bar in the drift region, and a reduced electric field of 2.7 kV/cm/bar in the EL region.}
\end{figure}

In addition, as a cross-check for the operation of our detector, we have also looked into the primary scintillation light produced by the interaction of \textsuperscript{241}Am alpha particles with the gas. Figure~\ref{fig:waveform} depicts a typical average waveform, obtained with the LeCroy WaveRunner 610Zi digital oscilloscope by averaging 2000 individual waveforms from alpha particle interactions in the GPSC volume, previously aligned to the instant when the EL amplitude reaches 50$\%$  of its maximum.

\begin{figure}[tbp]
\centering 
\includegraphics[width=13cm]{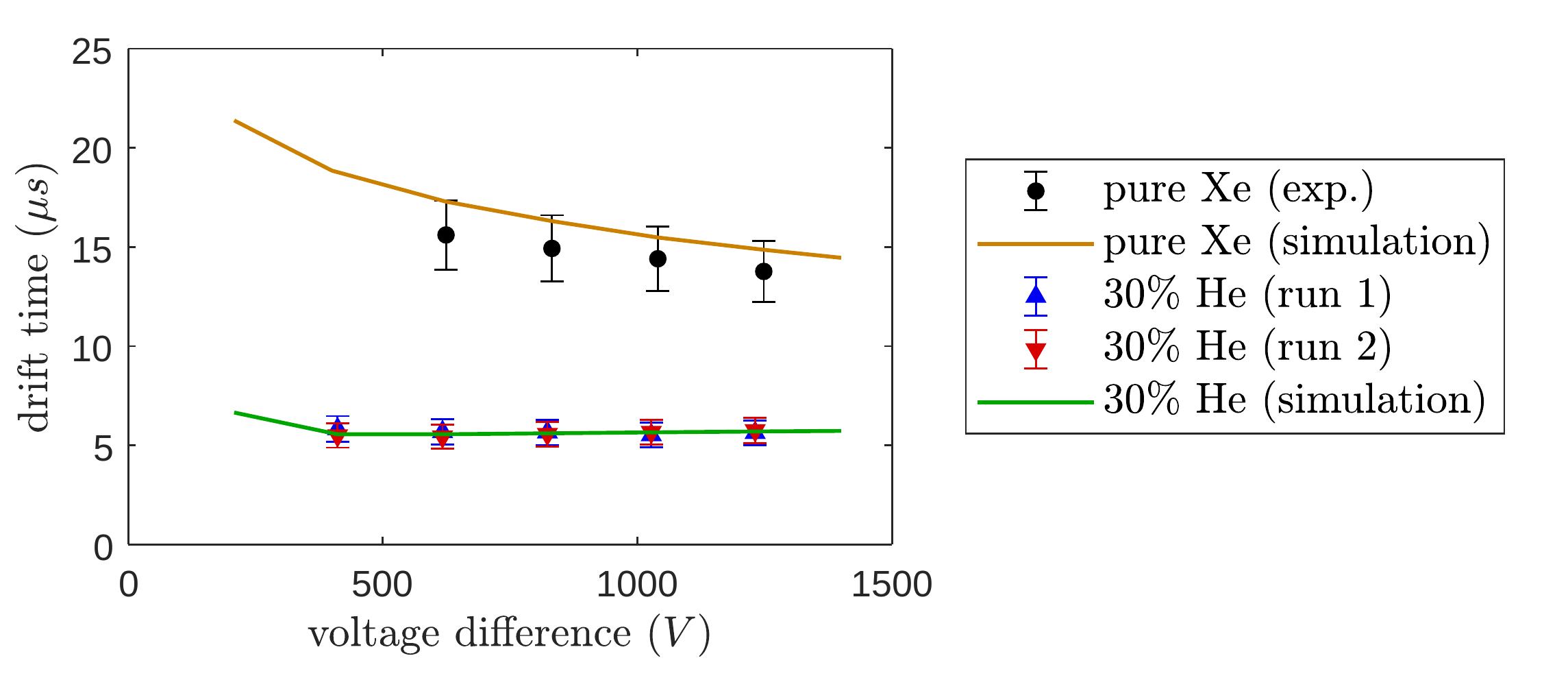}
\caption{\label{fig:vel}Electron drift times as a function of voltage differences applied to the drift region, for pure xenon and Xe-30$\%$ He.}
\end{figure}

The primary scintillation allows to determine the electron drift time while crossing the drift region and to compare it with the theoretical value obtained using the values of the drift electric field along the drift path and the values presented in the literature for the electron drift velocity in the gas. The results obtained with pure xenon and a 30$\%$  He mixture are shown in figure~\ref{fig:vel} for several voltage differences applied to the drift region. Two different series of measurements presented for the mixture of 30$\%$  He have been taken, with a time interval of seven days. The error bars show the systematic uncertainty for 90$\%$  C.L. and are mainly related to the calculations of the alpha particle penetration in the drift region. A difference lower than 10$\%$  with respect to the experimental values was found, showing a good agreement between experimental and simulation values within the errors. 

\begin{figure}[tpb]	
\centering
\includegraphics[width=7.4cm]{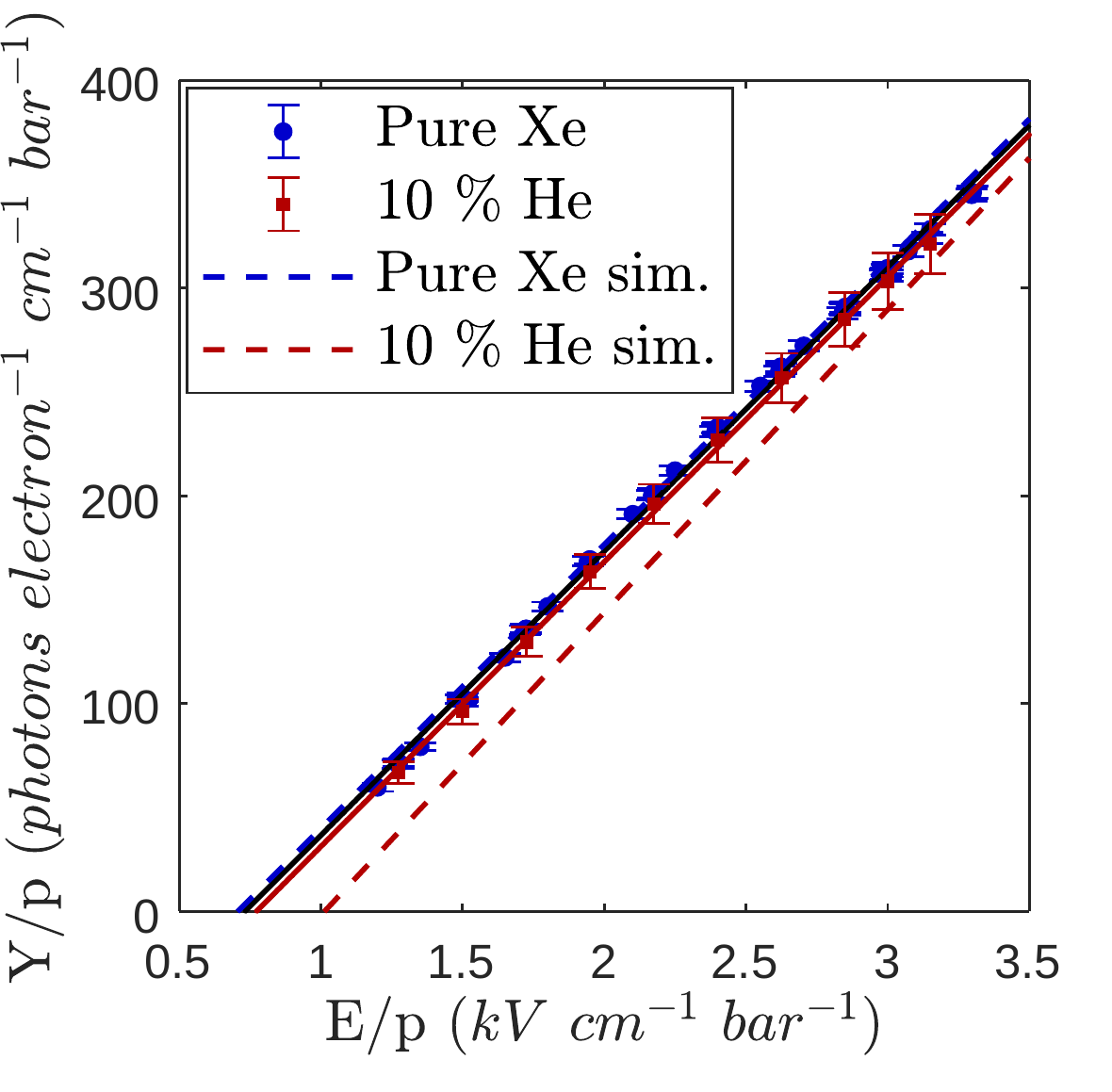}
\includegraphics[width=7.4cm]{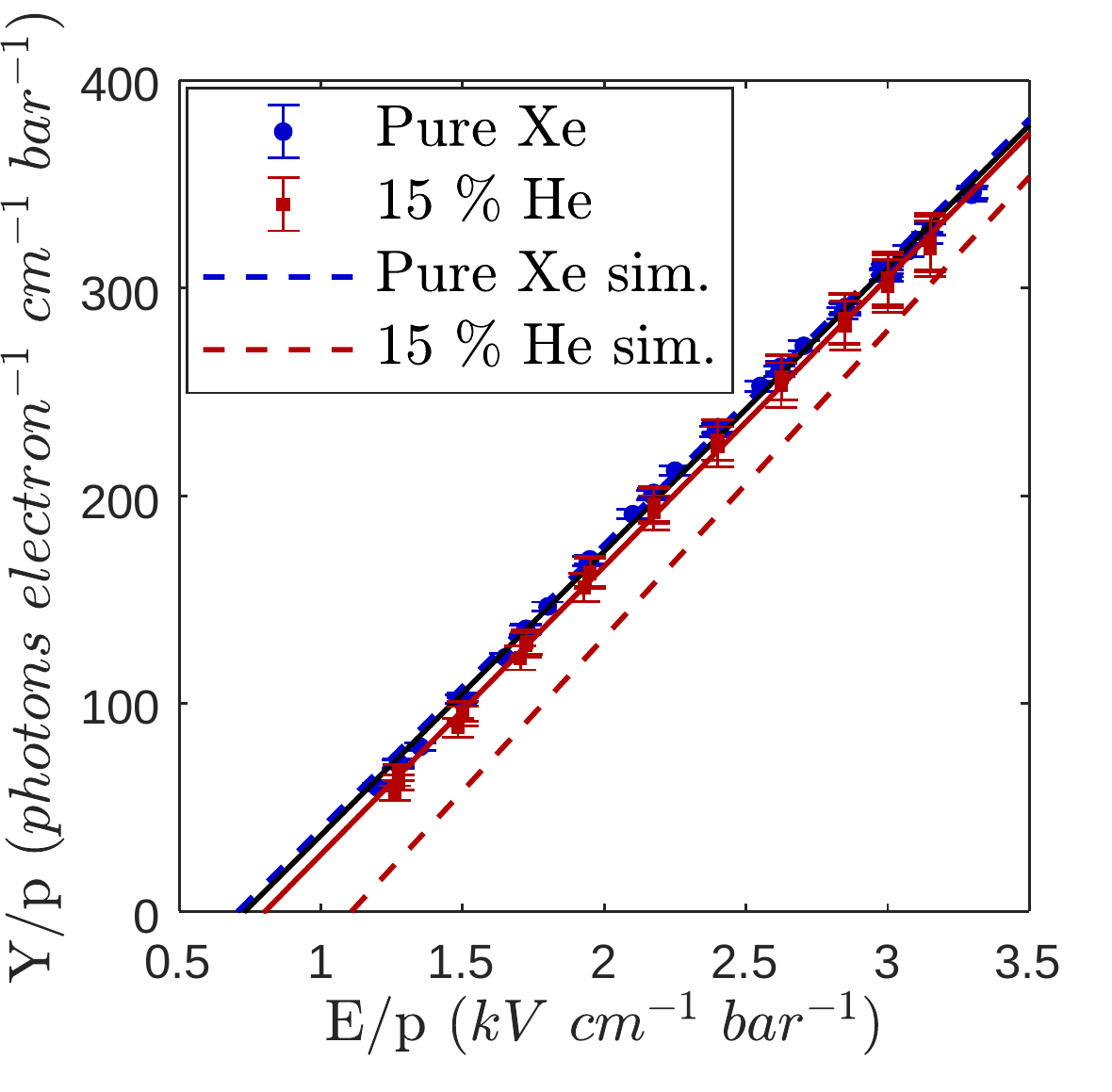}
\includegraphics[width=7.4cm]{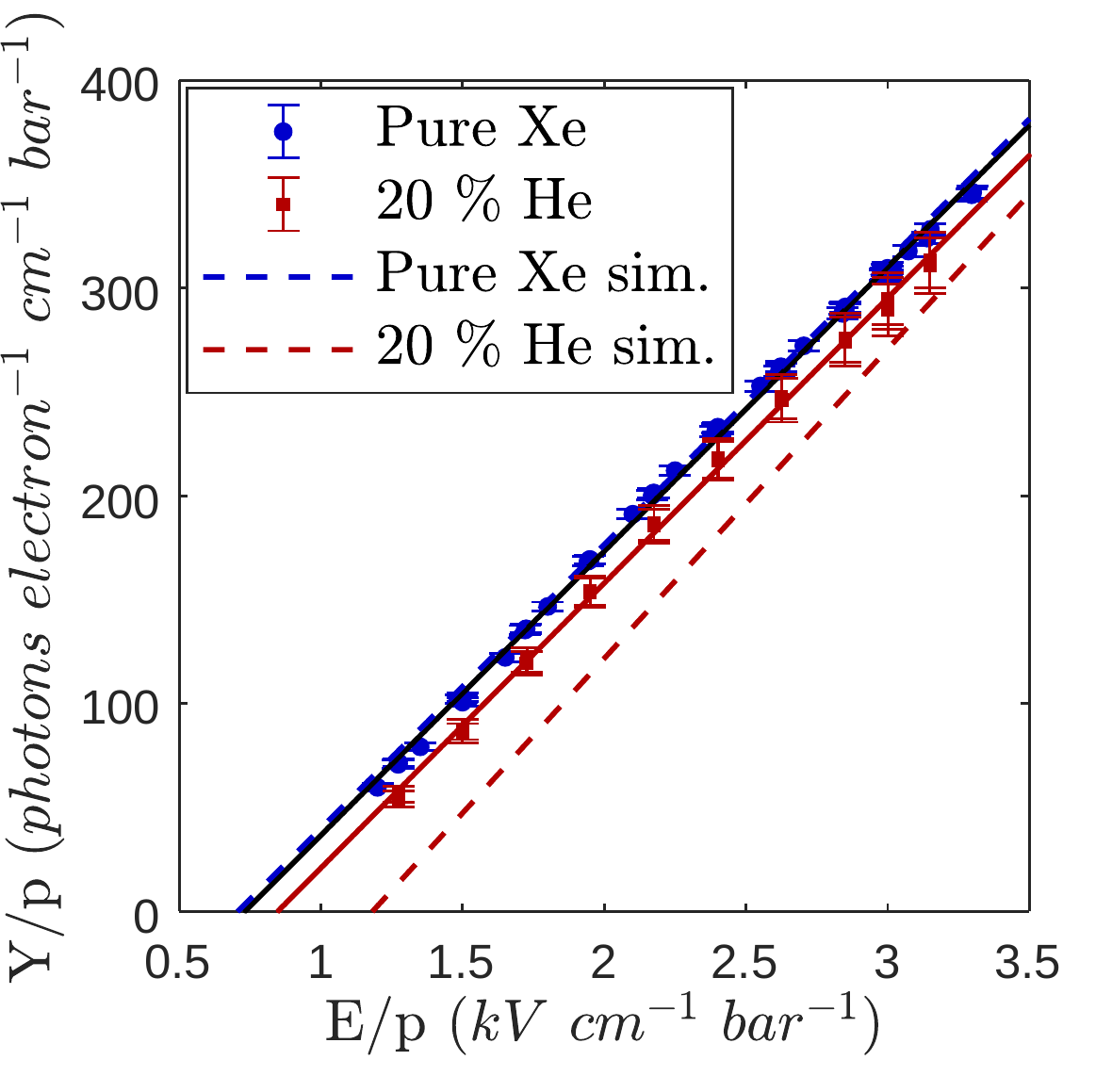}
\includegraphics[width=7.4cm]{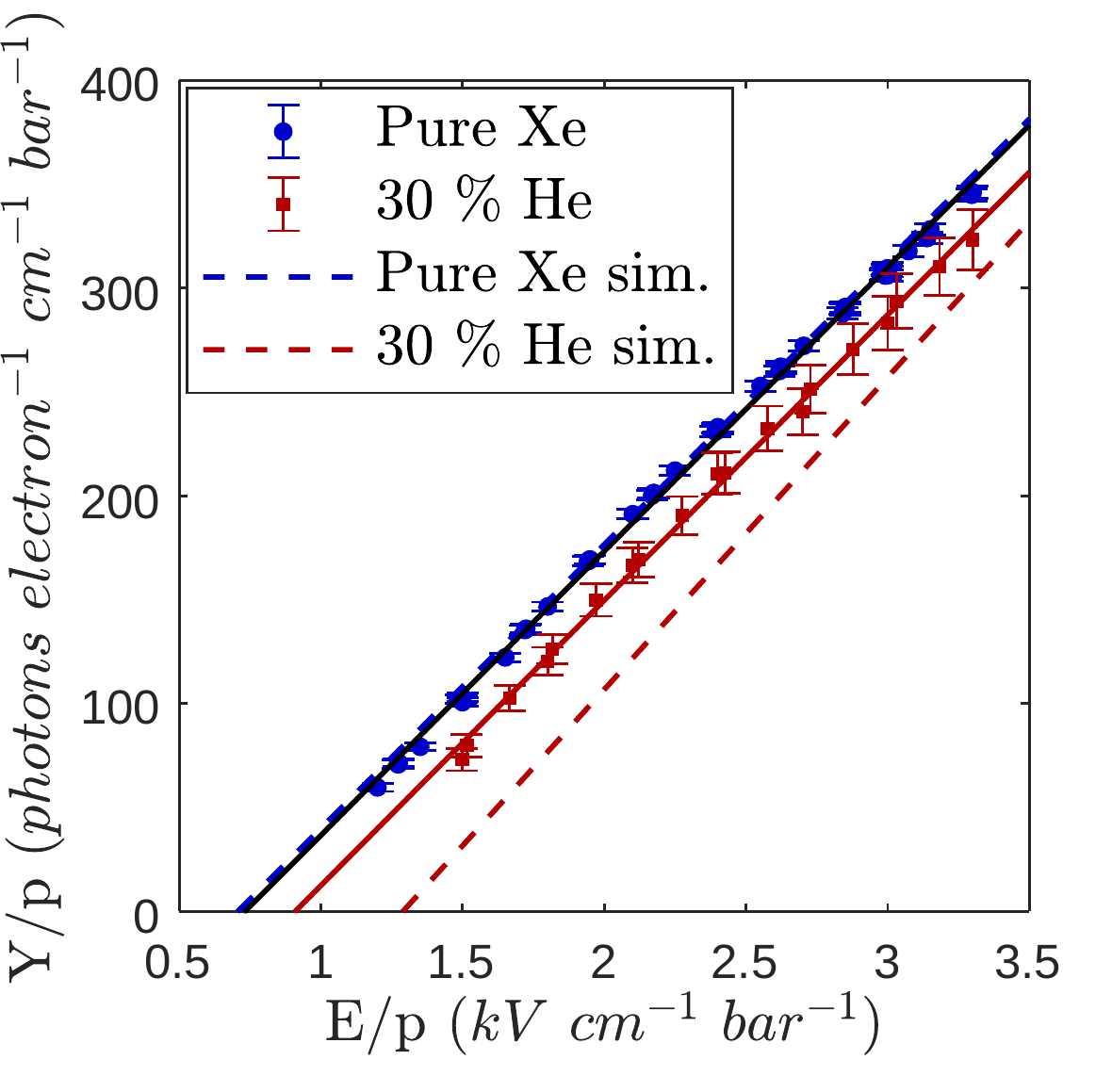}
\caption{\label{fig:Y_XeHe}Reduced electroluminescence yield as a function of reduced electric field in the scintillation region for pure xenon and the different Xe-He mixtures studied in this work. Solid lines show linear fits to the experimental data, while dashed lines are the simulation data obtained with the simulation tool of ref.~\cite{33}.}
\end{figure}

The EL yields of the studied Xe-He mixtures are presented in figure~\ref{fig:Y_XeHe} as a function of the reduced electric field applied to the scintillation region. Several mixtures have been made for each of the He-concentrations, namely three for 15$\%$  He, two for 20$\%$  He and only one for 10$\%$  He and for 30$\%$  He. Two different series of measurements are presented for the same mixture of 30$\%$  He, taken seven days apart. Along with the experimental data, figure~\ref{fig:Y_XeHe} shows the linear fits applied to the experimental data in each mixture (solid lines). For the mixtures of Xe-15$\%$ He and Xe-20$\%$ He, a single linear function was fitted to the whole set of data points, displaying the average linear trend for each He-concentration. For comparison, the simulation results obtained with the simulation tool of~\cite{33} are also depicted in figure~\ref{fig:Y_XeHe} (dashed lines). Table~\ref{tab:el_para} lists the EL amplification parameter and the scintillation threshold obtained from the linear fits to the experimental data for each of the studied mixtures. An additional systematic uncertainty of about 5$\%$  is estimated, being the main contribution due to the correction of the LAAPD leak current. 

The experimental values exhibit a lower reduction in EL than predicted by simulation. For instance, while the simulation results foresee a drop of $ \sim$16$\%$ in the EL yield of Xe-15$\%$ He at an E/p $ \sim$2.5 kV/cm/bar, when compared to the yield of pure xenon, in the experimental measurements this drop is only $ \sim$6$\%$. A possible contribution to this difference may be due to neutral bremsstrahlung, i.e. the bremsstrahlung emitted by electrons, scattered on neutral atoms, while drifting in the scintillation region~\cite{45}. This type of radiation might be extended from VUV to NIR~\cite{45}, a region where the APD is also sensitive. This issue has to be addressed in future studies.

\begin{table}[tbp]
\centering
\begin{tabular}{|c|c|c|}
\hline
He concentration&EL threshold&Amplification parameter\\
\hline 
0$\%$ &	0.73 $\pm$ 0.01 &	137 $\pm$ 1\\
10$\%$ &	0.77 $\pm$ 0.03 &	137 $\pm$ 2\\
15$\%$ &	0.80 $\pm $ 0.02 &	139 $\pm$ 1\\
20$\%$ &	0.85 $\pm $ 0.02 &	137 $\pm$ 1\\
30$\%$ &	0.91 $\pm $ 0.03 &	137 $\pm$ 2\\
\hline
\end{tabular}
\caption{\label{tab:el_para} Electroluminescence amplification parameter and scintillation threshold obtained from the linear fits to the experimental data for the studied mixtures.}
\end{table}

\begin{figure}[tbp]
\centering 
\includegraphics[width=12cm]{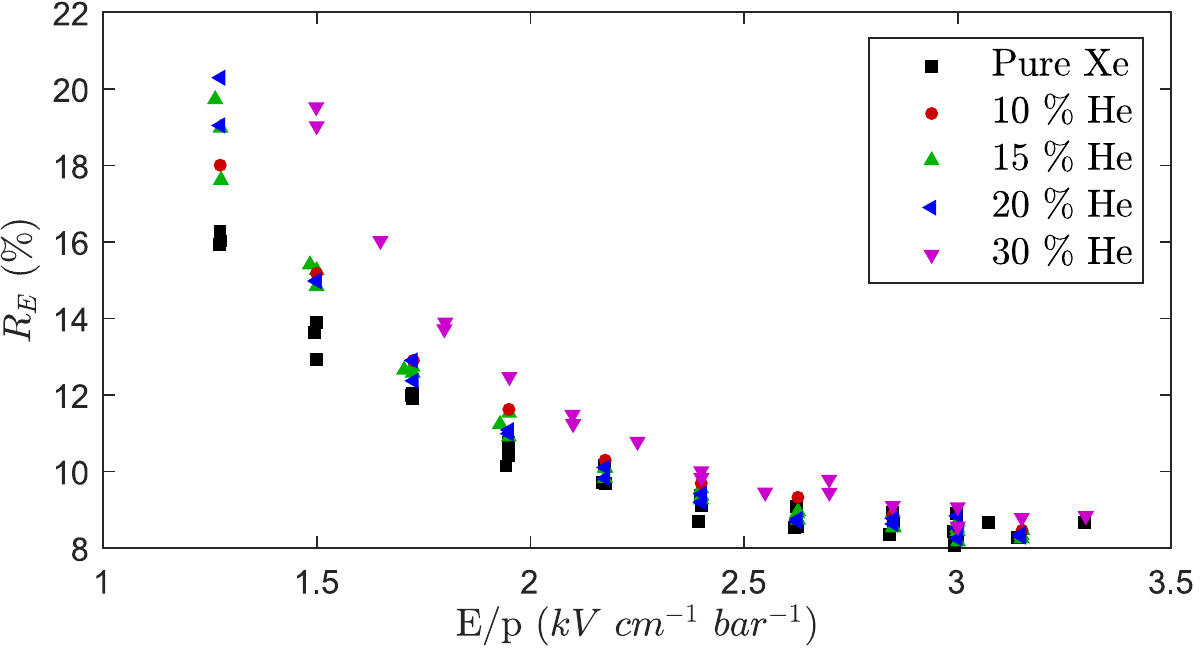}
\caption{\label{fig:R}Energy resolution as a function of reduced electric field in the scintillation region, obtained for pure xenon and the different Xe-He mixtures studied in this work. A relative statistical error of 3$\%$  was calculated for several measures having the same working conditions. }
\end{figure}

In figure~\ref{fig:R}, the GPSC energy resolution (FWHM) for the different pulse-height distributions is depicted as a function of reduced electric field in the scintillation region, for the Xe-He mixtures studied in this work. Within experimental uncertainties, no significant differences are perceived in the values of the achieved energy resolution for the different Xe-He mixtures, for E/p values above 2.0 kV/cm/bar. 

The experimental results of figures~\ref{fig:Y_XeHe} and~\ref{fig:R} show that helium addition to xenon in the 0-30$\%$  range does not significantly reduce the EL yield and the associated statistical fluctuations, as already foreseen from simulation results. Therefore, concerning the EL yield, helium is a much better option to be used as additive to pure xenon, in optical TPCs, than molecular additives. 

On the other hand, helium addition results in the reduction of the overall \textsuperscript{136}Xe mass which, per se, contributes to the reduction of the TPC sensitivity to DBD detection. The sensitivity to m\textsubscript{$\beta \beta$}, the so-called effective Majorana mass of the electron neutrino, of an experiment searching for 0$ \nu  \beta  \beta $  decay, i.e. the m\textsubscript{$ \beta \beta $} lower limit that an experiment may achieve, is given by~\cite{4}
\begin{equation}
\label{eq:sens}
S(m_{ \beta  \beta }) = k \; \sqrt[]{ \frac{\overline{N}}{ \varepsilon Mt}} ,
\end{equation}
where $k$ is a constant, $\overline{ N }$ is the average upper limit of the number of observed events expected in the experiment under the no-signal hypothesis, $ \varepsilon $ is the signal detection efficiency, M is the source mass and $t$ is the measuring time.

In the presence of a dominant $ \beta \beta $2$\nu$ background, the average upper limit for $\overline{ N }$  is proportional to the square root of the mean number of background events, i.e. $\overline{ N }\propto  \sqrt{b}$. Moreover, the number of background events is usually proportional to the exposure, $M \cdot t$, and to the width of the energy window defined by the resolution of the detector, $ \Delta E$, i.e. $ b = c \cdot M \cdot t \cdot \Delta E$, where c is the expected background rate. Therefore, the sensitivity becomes dependent on~\cite{4}:

\begin{equation}
\label{eq:sens2}
S(m_{ \beta  \beta }) \propto \frac{ \Delta E^{1/4}}{M^{1/4}} .
\end{equation}

Since there is no degradation in the TPC energy resolution, as demonstrated in the present studies, a 15$\%$ reduction of the \textsuperscript{136}Xe mass will result, per se, in a sensitivity degradation of $ \sim $4.1$\%$ in the TPC sensitivity. Nevertheless, although the reduction in the target mass implies a reduction effect on the sensitivity, the improvement of the electron diffusion and, ultimately, on the topological discrimination efficiency to background will enable a more sensitive search for DBD. In addition, the resulting increase in the electron drift velocity will have a positive reduction in the electron attachment to impurities, which are non-trivially distributed throughout the detector, and which adds some space- and time-dependent fluctuations to the charge yields as well as to charge loss.

The above reduction in sensitivity is to be compared with the case of Xe-0.15$\%$  CH\textsubscript{4} mixture~\cite{31}, assuming that a similar background suppression can be achieved for both types of low-diffusion mixtures. For the Xe-0.15$\%$  CH\textsubscript{4} mixture, the variation in the TPC \textsuperscript{136}Xe mass is negligible, while a small degradation of the energy resolution results in a $ \sim $1$\%$ and $ \sim $3$\%$ reduction of the TPC sensitivity for a light collection efficiency of 3$\%$ and 0.5$\%$, respectively, and considering an additional constant contribution of 0.5$\%$ to the overall energy resolution in the NEXT TPC. However, other practical aspects such as the impact of CH\textsubscript{4} quenching on the primary scintillation signal and the long term purification and stability are factors that have to be considered as well. 

These issues are to be investigated in larger TPC prototypes such as NEXT-DEMO and/or NEXT-NEW in subsequent R$\&$D programs. A direct measurement of the electron transverse diffusion is still pending as well as the pressure scaling assumption that the 1 bar measurements can extrapolate to 10 bar at the same E/P.

\section{Conclusions}
\label{sec:conc}

In this paper we experimentally confirm that the addition of helium to pure xenon in the concentration range of 0-30$\%$  does not reduce significantly the electroluminescence yield of the resulting mixture. For a typical reduced electric field of 2.5 kV/cm/bar in the scintillation region, the EL yield is reduced by $ \sim $2$\%$ , 3$\%$ , 6$\%$  and 10$\%$  for 10$\%$ , 15$\%$ , 20$\%$  and 30$\%$  He concentration, respectively. No degradation was observed in the detector energy resolution with the addition of helium to pure xenon. The EL yield decrease is less than what has been obtained from the most recent simulation framework in the literature. The present results are an important benchmark for the simulation tools to be applied to future optical TPCs based on Xe-He mixtures.

It is noted that the impact of the helium addition is lower than that expected from the simulation results of~\cite{33} where, e.g., a reduction of $ \sim $12$\%$  is foreseen for the Xe-15$\%$ He mixture at a reduced electric field of 2.5 kV/cm/bar in the scintillation region. These results, combined with those obtained for the drift-diffusion properties in the range of 1-10bar~\cite{34}, reinforce the potential of Xe-He admixtures for $\beta  \beta $ searches. 

\acknowledgments

The NEXT Collaboration acknowledges support from the following agencies and institutions: the European Research Council (ERC) under the Advanced Grant 339787- NEXT; the European Union’s Framework Programme for Research and Innovation Horizon 2020 (2014-2020) under the Marie Sklodowska-Curie Grant Agreements No. 674896, 690575 and 740055; the Ministerio de Economía y Competitividad of Spain under grants FIS2014-53371-C04, RTI2018-095979, the Severo Ochoa Program SEV-2014-0398 and the María de Maetzu Program MDM-2016-0692; the GVA of Spain under grants PROMETEO/2016/120 and SEJI/2017/011; the Portuguese FCT under project PTDC/FIS-NUC/2525/2014, under project UID/FIS/04559/2013 to fund the activities of LIBPhys, and under grants PD/BD/105921/2014, SFRH/BPD/109180/2015; the U.S. Department of Energy under contracts number DEAC02-06CH11357 (Argonne National Laboratory), DE-AC02-07CH11359 (Fermi National Accelerator Laboratory), DE-FG02-13ER42020 (Texas A$\&$ M) and DE-SC0019223 / DESC0019054 (University of Texas at Arlington); and the University of Texas at Arlington. DGD acknowledges Ramon y Cajal program (Spain) under contract number RYC-2015-18820. We also warmly acknowledge the Laboratori Nazionali del Gran Sasso (LNGS) and the Dark Side collaboration for their help with TPB coating of various parts of the NEXT-White TPC. Finally, we are grateful to the Laboratorio Subterraneo de Canfranc for hosting and supporting the NEXT experiment.


\end{document}